\documentclass{article}

\usepackage[a4paper]{geometry}

\usepackage{amsmath,amssymb}

\usepackage{graphicx}

\newtheorem{definition}{Definition}

\title{\textbf{Sustainable Credit And Interest Rates}\footnotemark[1] \footnotetext[1] {This research has been done at Dublin City University. The author gratefully acknowledges the Science Foundation Ireland (Edgeworth Center and FMC2) for their support.}}

\author{\small{\textbf{Andreas Hula}}\\ \footnotesize{School Of Mathematics, Dublin City University,Dublin 9,Dublin,Ireland}\\ \footnotesize{(andreas.hula@dcu.ie)}}

\date{\today}

\begin{document}
\maketitle
\tableofcontents
\section{Introduction}
With reduced or even negative growth in the goods and services economy in many countries and debt levels which become an increasing burden on developed societies, the calls for a change in economic policy and even the monetary system become louder and increasingly impatient.\\
We research the consequences of a system of credit and debt, that still allows for the expansion of credit and fundamentally retains many features of the present monetary system, without the instability inherent in the present system. Many alternative systems have been proposed. Our approach rest on the concept of productivity share promises. The motivating example in $(\ref{Mot})$ and its issues give rise to the concept of the productivity numeraire $(\ref{ProdNum})$. We will describe the implementation we have in mind in the following sections $(\ref{Cred})$ and $(\ref{Def})$. We extend the concept to services $(\ref{paidservices})$ and to private credits $(\ref{PrivCred})$. In section $(\ref{GovBond})$ we explain how government bonds would be issued in this system. Briefly we discuss how interbank lending would have to be handled to prevent instabilities in $(\ref{IBL})$. We then move on to discuss the effect such a model might have on the term structure of interest rates, represented by the HJM equation $(\ref{Forward})$. The relation to the HJM condition allows to formulate a no arbitrage condition based on an economic growth model $(\ref{noab})$ . The implications of our approach are summarized in $(\ref{Implications})$ before we reach our conclusions in $(\ref{conclusions})$.\\
A good source for the mathematics used here is the book by Bj\"ork $\cite{BJORK}$.
\section{Motivation}
\label{Mot}
A historical way of handling interest based on productivity is the following:\\
A creditor lends the debtor 10 units of currency. The debtor produces goods worth those 10 units of currency to repay the loan and additionally goods worth a further unit of currency for interest. These goods (sold or not) belong to the creditor who has been repaid and received interest.\\
Of course the problem here is, that the future market value $S_t$ of the produced goods (let us count them in a process $N_t$) is unknown. Therefore it may become impossible to even repay the original loan.\\
For that reason, one might look for a way of bringing promises more into line with what can reasonably well be predicted about the future.
\section{The Productivity Numeraire}
\label{ProdNum}
Given an activity to produce objects or deliver services, we can measure the productivity of the activity in terms of the number of goods produced or the number of services delivered. This will then be a positive process (depending on the way this is measured it may not be integer valued) and for reasonable modeling assumptions a semimartingale. We denote this numeraire as $(N_t)_t$.\\
Let $\Pi_{\frac m n}$ denote the productivity measured in $N_t$ .
\section{Credit On Material Productivity}
\label{Cred}
The most central element of a sustainable way of giving credit, is to tie the interest to the productivity of the enterprise financed by the credit. The simplest example is a credit to an enterprise where goods are to be produced. \\
The successful borrower will lead the enterprise, while the bank finances the cost of setting up the enterprise and gets to check (but not decide) on the practice of the enterprise if it desires.\\
Let the cost of the enterprise to be setup be $C$ (let environmental costs be included in this) in the currency of the 2 partners.\\
We want to think of the enterprise in terms of money and in terms of productivity, so we need $2$ numeraires: $N_t$ as described above and $S_t$ as price process of the produced goods and services. We may assume any type of model for the evolution of the change of numeraire $(\frac{N_t}{S_t})_t$, which is just the market price of the amount of goods (or services) $N$. That allows to calculate the expected repayment rate at rate $r_t$:
\begin{equation}
rpr_{\frac{m}{n} T}=\frac{1}{\mathbb E [\Pi_{\frac {m}{n}T}-\Pi_{{\frac {m-1}{n}T}}]}\mathbb E (\frac{ C }{n }\frac{N_{\frac m n T}}{S_{\frac m  n T}}).
\end{equation}
Here $T$ is the time horizon for repayment and $n$ is the number of time steps in which the repayment is to occur. Thus $rpr_t$ is the expected part of the production that will allow us to repay the given sum $C$ in $n$ total steps. The expectation is fixed at time $0$, so the producer knows the exact number at the time the loan is agreed on and is not subject to "unexpected" price developments of his enterprise $\frac{S_t}{N_t}$. That risk has to be carried by the bank and it might result in unexpected gains as well as losses.\\
 A total sum of 
\begin{equation}
\Pi = \sum_{m=1}^n  rpr_{\frac{m}{n} T} (\Pi_{\frac{m}{n} T}-\Pi_{{\frac {m-1}{n}T}})
\end{equation}
in productive results has been accrued to cover just the loan. The market value depends on the agreed on selling procedure but will typically be
\begin{equation}
A =\sum_{m=1}^n rpr_{\frac{m}{n} T} (\Pi_{\frac{m}{n} T}-\Pi_{{\frac {m-1}{n}T}}) \frac{S_{\frac m n T}}{N_{\frac m n T}}.
\end{equation}
meaning the product is always sold on the market right away. (Other options may be considered in the future).\\
An additional interest paying period $I$ with $m$ payments will generally be agreed on. The interest fraction $rI$ should be considerably lower than the repayment fraction $rpr$. The total interest will then be
\begin{equation}
J = \sum_{m=1}^n rI_{\frac{m}{n}I} ( \Pi_{\frac{m}{n}I}-\Pi_{{\frac {m-1}{n}I}}) \frac{S_{\frac m n I}}{N_{\frac m n I}}.
\end{equation}
Thus, the total sum repayed for the investment $C$ will be
\begin{equation}
E = A +J.
\end{equation}
\section{Default And Loss}
\label{Def} 
What happens, if the sum $C$ is bigger than $E$? Obviously the bank has financed more than the fraction of the productive results of the enterprise could repay when sold on the market. In that case, as with any bad bet, the bank has to accept the losses. There can be no ownership of productive assets or any enterprises by the bank. \\
The bank will take a loss of
\begin{equation}
C-E.
\end{equation}
That loss has to be covered by the banks wealth. \\
A more difficult situation occurs, if the enterprise fails altogether and production stops or the enterprise does not abide by good standards of conduct (diverts funds, willfully stops producing etc). In that case, the bank will have to deal with the local authorities. If there is an interest to continue the enterprise this can be organized locally and the loan continues, possibly with an negotiated extension.\\
 If there is no interest by the local authorities to continue the enterprise the bank has 2 more options. It can suggest a private investor which can continue the enterprise and the loan if the local authorities agree. Otherwise it must dismantle every unused element of the enterprise and every reusable element of the enterprise must be sold by the bank at the market and the obtained sum will be credited to the bank. All other elements (including the costs of dismantling) will have to be paid for by the bank. \\
How much may the bank lend? Basically, we assume a fractional lending system remains in place. But the lent money will have to show up in the banks balance. The total ratio of given credit to deposits for any given bank must be official and every depositor has to be informed regularly of this ratio.\\
\section{Credit For (Paid) Services}
\label{paidservices}
Of course we want to fund services as well. It is sometimes hard to find a market value for those, but if they are being paid for, we can formulate the loan in terms of hourly pay as productivity. Other than that everything applies as above with the qualification that there will usually not be as many possibilities to obtain money from a failed service enterprise as from a production enterprise. 
\section{Credit For Private Persons}
\label{PrivCred}
Of course things get more complicated when extending credit to private customers, who might for instance want to build a house or finance higher education. The basic principle must stay the same: Payments can only come from real produced wealth and a bank must not own real assets. Directly or indirectly.\\
Thus, for the housing example, we can take the present income. The bank is promised a fixed fraction of the borrowers income, working or otherwise. The borrower promises to keep the agreed on income sources flowing (usually this will be a pledge to work), otherwise the case will be considered to be a default. Legal provisions as to when this is the case will be crucial. A normal job loss will certainly not qualify as a default.\\
Apart from that, whatever the income development of the borrower, the bank can not claim more money than the agreed on fraction and has no rights over the house or the way of education chosen by the borrower, except in a willful default.\\
The education example could be an agreement on a start of repayments after the education.\\
The implications of the obligations on the borrower are thus much easier to understand for the borrower, making it much easier to reach a reasonable decision on whether to take up a loan or not.
\section{Government Bonds}
\label{GovBond} 
In such a system, it would be much more natural for the government to issue bonds on a fraction of its future economic output. A zero coupon bond, promising the buyer $1$ percent of the states tax revenue over the time span $[t,T]$ would be worth
\begin{equation}
B_p(t,T)= \frac 1 {100} \tau (t) e^{(\int_t^T f_p(t,s) ds )}
\end{equation}
where $B_p(t,T)$ is the price of the bond, $(\tau(t))_t$ is the state tax income process and $f_p(t,T)$ is the growth rate of tax revenue (which may be negative) of the tax generating economy. The big advantage being, that this relates the bond to real tax income and the rate $f_p(t,T)$ can be observed empirically, through the tax intake.\\
In terms of expectations of the lenders we can define 
\begin{equation}
\Gamma (t,T)=\frac 1{\mathbb E_{Lenders} ( \theta (t,T)))}.
\end{equation}
Differently said, this is the share of state income over the time span $[t,T]$ the lenders demand to owe $1$ unit of currency and thus a good measure of trust in the states economy.\\
\section{Inter Bank Lending}
\label{IBL}
A loan between $2$ banks is determined as with any other enterprise. There is an expectation on the performance of the banks projects and then money can be lend to the usual terms of repayment fractions and interest.\\
However there is one big extra provision: that money must not be used as deposit to base new loans on, but only as credit sum to directly fund projects. Liability is first in the hands of the receiving bank. If that bank collapses the lending bank is fully liable. Deposits must not be transfered without direct consent by the account holders at any given time (no general consent is possible). They must never be used to issue loans in other banking institutes.
\section{The Heath-Jarrow-Morton Analogue In A Sustainable Credit World}
\label{Forward}
The HJM equation for the forward rate process $f(t,T)$ is given through 
\begin{equation}
df(t,T)=\alpha (t,T)dt + \sigma(t,T)^T dW^*_t+ \int_{\mathbb{R}^r} \delta (t,x,T) (\mu - \nu^*)(dt,dx)
\end{equation}
See Appendix for a detailed explanation  The form of the equation needs not change. To see what happens to the forward rate, it is best to look at bonds issued by the country in terms of productive promises. The yield here is truly "default free" apart from the unavoidable political risk (free decision to default without need, massive revolution).\\
The empirically charged bond prices actually reflect the assertion of markets of the countries capability to run a healthy productive economy, instead of its ability to service ever growing interest.\\
We get the following equation from the bond dynamics
\begin{equation}
f_p(t,T)=\frac{\partial}{\partial T}\log (B_p (t,T)) 
\end{equation}
thus the productive state economy forward rate would be equal to the negative growth rate of the tax income of the state.\\
We assume this forward rate to be described by the HJM framework
\begin{equation}
df_p(t,T)=\alpha_p (t,T)dt + \sigma_p(t,T)^T dW^*_t+ \int_{\mathbb{R}^r} \delta_p (t,x,T) (\mu - \nu^*)(dt,dx)
\end{equation}
so the condition for no arbitrage, given a certain dynamic model would be
\begin{equation}
A_p(t,T)+\frac12 |S_p(t,T)|^2 +\int_{\mathbb{R}^r} (e^{D_p(t,xT)}-1-D_p(t,x,T))F(t,dx)=0\qquad [d\mathbb{P}^*\times dt]
\end{equation}
excluding existence issues for the moment.\\
If we assume some economic growth model to describe the drift of $r_p(t)$, we can tie the financial concept of the forward rate dynamics to economical growth theories.\\
We can also use the "lenders demand process" $\Gamma (t,T)$ to define
\begin{equation}
f_{\Gamma} (t,T)=\frac{d}{dt}\log \Gamma (t,T).
\end{equation}
The properties of this rate will be a subject of further study.\\ 
Those equations are infinite dimensional SDEs much like the original HJM equation. In fact, we may model them in much the same way. 
\section{No Arbitrage In Sustainable Credit}
\label{noab}
Using a HJM framework for one of the related forward rates above allows to apply the HJM framework to determine no arbitrage conditions on the forward rate. Such conditions can connect economic and financial points of view, if we use economic models to describe $r_p(t)$ and then check the HJM drift condition based on those models. \\
Assume from some growth model
\begin{equation}
d \alpha_p(t,T) = g(t,T,\alpha_p(t,T))
\end{equation} 
then the drift condition in the HJM framework implies
\begin{equation}
\int_t^T \alpha_p (t,s) ds =-\frac12 |S_p(t,T)|^2 -\int_{\mathbb{R}^r} (e^{D_p(t,xT)}-1-D_p(t,x,T))F(t,dx)\qquad [d\mathbb{P}^*\times dt]
\end{equation}
and we can calculate for instance $\frac 12|S(t,T)|^2$ for a continuous diffusion model in mathematical finance, by using the drift taken from an economic model.\\
It is important to note, that this could be applied even in the present system, provided we could define a model to give us an analog of $f_p(t,T)$.
\section{Implications Of The Model}
\label{Implications}
The concept of interest based on productive activity has the advantage, that in terms of production shares there can be no unavoidable default for a state.\\
 It also ensures, that a states production will not end up entirely serving interest payments. Thus the rates are first of all truly default free except for the political risk.\\
Any obligations in this model wear off after the agreed on time. Debt therefore grants less political influence over state policies and private debtors. Withholding of credit of course would still be an option, but both public pressure and the founding of new banks can likely prevent that measure from influencing policy as much as is presently the case.\\
However the model is a true market model, in that it leaves both risk and reward primarily with the investor/creditor while making "too big to fail" type situations far less likely.\\
Furthermore, this model could provide a smooth transition between the present monetary system and others, by allowing the state to become debt free or almost debt free in a reasonable time frame. 
\section{Conclusions}
\label{conclusions}
\begin{itemize}
\item
The model prevents debt ratios above $100$ percent of the state income and makes sure the agreed on debt is always repayable.
\item
The model can be used both, in a debt based monetary system or in a system where the state can print freely.
\item
The total amount of money will generally expand if there are productive activities but may contract without causing a economic collapse.
\item
The HJM framework can directly link economical models of growth to financial models for the productive forward rate and vice versa.    
\end{itemize}
\appendix
\section{HJM Framework}
The HJM equation is given through 
\begin{definition}[HJM Framework]
The following exposition is taken from $\cite{EO2005}$ and gives a very general framework.\\
We assume a complete stochastic basis $(\Omega,\mathcal F,\mathbf F,\mathbb P^*)$ satisfying the usual conditions. Then we assume the following dynamics for the instantaneous forward rate
\begin{equation}\label{HJMEQ}
df(t,T)=\alpha (t,T)dt + \sigma(t,T)^T dW^*_t+ \int_{\mathbb{R}^r} \delta (t,x,T) (\mu - \nu^*)(dt,dx)
\end{equation}
where $W^*$ is a standard Brownian Motion in $\mathbb{R}^d$, $\mu$ is the random measure of jumps of a semimartingale with continuous compensator $\nu^*$ for which $\nu^*(dt,dx)=F(t,dx)dt$ is assumed to hold. The coefficients are continuous in the second variable. $\alpha:\Omega \times [0,T^*] \times [0,T^*]\rightarrow \mathbb{R}$ and $\sigma:\Omega \times [0,T^*] \times [0,T^*]\rightarrow \mathbb{R}_+^d$ are assumed to be $\mathcal{P} \times \mathcal{B} ([0,T^*])$ measurable and $\delta:\Omega \times [0,T^*]\times \mathbb{R}^r\times [0,T^*]\rightarrow \mathbb{R}$ is assumed to be $\mathcal{P}\times \mathcal{B}(\mathbb{R}^r)\times \mathcal{B}([0,T^*])$ measurable. We denote $\Delta := \{(s,u)\in\mathbb{R}_+\times\mathbb{R}_+| 0\leq s \leq u \leq T^* \}$. Then
\begin{itemize}
\item
If $(t,T)\notin \Delta$ we have $\alpha(t,T)=\delta(t,x,T)=0$ and $\sigma(t,T)=(0,0,\dots,0)^T$.
\item
For all $(t,T)\in\Delta$ there holds
\begin{equation}
\int_0^T \int_t^T |\alpha(s,u)|duds < \infty
\end{equation}
\begin{equation}
\int_0^T \int_t^T |\sigma (s,u)|^2 duds < \infty
\end{equation}
\begin{equation}
\int_0^T \int_{\mathbb{R}}\int_t^T |\delta (s,x,u)|^2 du \nu^*(ds,dx).
\end{equation}
\item
We denote
\[
A(t,T):= -\int_t^T \alpha (t,u)du\qquad S(t,T):=-\int_t^T \sigma (t,u)du
\]
\[
D(t,x,T)=-\int_t^T \delta(t,x,u)du
\]
\item
We get two conditions for $\mathbb{P}^*$ to be a martingale measure. The first
\begin{equation}
\int_0^t \int_{\mathbb{R}^r} e^{D(s,x,T)}-1-D(s,x,T) F(s,dx)ds< \infty \qquad \forall (t,T)\in\Delta
\end{equation}
\item
The second
\begin{equation}
A(t,T)+\frac12 |S(t,T)|^2 +\int_{\mathbb{R}^r} (e^{D(t,xT)}-1-D(t,x,T))F(t,dx)=0\qquad [d\mathbb{P}^*\times dt]
\end{equation}
\end{itemize}
\end{definition}
\newpage

\bibliography{dissertatio}

\vfill
\end{document}